\documentclass[aps,twocolumn,superscriptaddress,floatfix,longbibliography]{revtex4-1}

\usepackage{bm,amsmath,amsfonts,amssymb,braket}
\usepackage{latexsym,dsfont,array,layout,mathrsfs,textcase}
\usepackage{graphicx,graphics,subfigure,xcolor}
\usepackage{comment,verbatim}
\usepackage{float}
\usepackage{times}
\usepackage{xspace}
\usepackage{stmaryrd}
\usepackage{multirow}
\usepackage{mathtools}
\usepackage{booktabs}
\usepackage{CJK}
\usepackage[colorlinks,linkcolor=blue,citecolor=blue,urlcolor=blue]{hyperref}

\begin{document}
\title{Dual-Space Invariance as a Universal Criterion for Multifractal Critical States}

\author{Tong Liu}
\thanks{t6tong@njupt.edu.cn}
\affiliation{School of Science, Nanjing University of Posts and Telecommunications, Nanjing 210003, China}

\date{\today}

\begin{abstract}

In Anderson localization, eigenstates of disordered quantum systems are broadly classified as extended, localized, or critical. Although critical states exhibit multifractal character, a precise and operational criterion for their identification remains an open challenge, as Lyapunov exponents in real space cannot uniquely distinguish them from extended states. 
Here we address this challenge by asserting that critical states are uniquely characterized by an emergent dual-space invariance between position and momentum space. Building on the Liu--Xia criterion of the simultaneous vanishing of Lyapunov exponents ($\gamma=\gamma_m=0$), we show that this dual-space invariance extends beyond Lyapunov exponents and governs wavefunction scaling, revealing a fundamental property inaccessible from either space alone. 
Through numerical simulations, we demonstrate that the inverse participation ratio exhibits matching scaling behavior in position and momentum space for critical states, in sharp contrast to extended and localized states, which display a pronounced asymmetry between the two spaces. 
This dual-space invariance provides a direct, robust, and universal criterion for identifying multifractal critical states. Our results establish a fundamental principle of Anderson criticality and open new avenues for its detection in modern quantum simulation platforms.

\end{abstract}
\maketitle

\section{Introduction}

Anderson localization~\cite{photonic,Critical0} is a fundamental manifestation
of wave interference in disordered media, leading to the suppression of
diffusive transport~\cite{Liu-self-duality,Li1,Li2}.
Within such systems, single-particle eigenstates are commonly classified as
extended (metallic), localized (insulating), or critical.
Critical states--also referred to as multifractal states--occupy an
intermediate regime between extended and localized phases~\cite{You,g1,zhou}.
Unlike extended states, which exhibit nearly uniform probability
distributions, or localized states, which decay exponentially, critical
states typically display power-law spatial decay~\cite{Zhang,q2}.
Their hallmark features include multifractality, characterized by a
continuous spectrum of anomalous scaling exponents~\cite{q1,q4,gg10,gg11}, as
well as scale invariance and statistical self-similarity arising from the
competition between localization and delocalization~\cite{gg12,q5,q6}.
In paradigmatic settings such as the three-dimensional Anderson model, critical states occur at mobility edges separating extended and localized regimes~\cite{NH3,Yao1,Yao2}.

Quasiperiodic systems~\cite{cai,Slager,SSHQ,gg3,Ayan1,Ayan2} provide an alternative platform for realizing critical states, even in the absence of true
randomnes.
Recent progress in critical states has uncovered a variety of
nontrivial localization phenomena, including anomalous mobility
edges~\cite{Liu-Anomalous}, renormalization-group descriptions~\cite{RG1,RG2},
recursion-defined real spectra~\cite{Liu-Real1,Liu-Real2}, and
coupled-chain-induced criticality~\cite{twochain}.
In previous theoretical descriptions~\cite{Jiang1,Jiang2,gg5,gg6}, the asymptotic localization properties of single-particle eigenstates were typically characterized by a single real-space Lyapunov exponent $\gamma$. The Lyapunov exponent, as a dynamical indicator, signals the localization-delocalization transition.
For exponentially localized states, the wave-function amplitudes decay
as $|\psi_n| \propto e^{-\gamma |n-n_0|}$,
where $n_0$ denotes the localization center and $\gamma>0$ quantifies the
inverse localization length~\cite{gg9,Duncan}.
A finite Lyapunov exponent $\gamma>0$ signals exponential localization,
whereas $\gamma=0$ indicates delocalization.
It is important to emphasize, however, that a vanishing Lyapunov exponent $\gamma=0$ in real space does not uniquely characterize extended states: both truly extended states and critical (multifractal) states fall into the non-exponentially localized class and exhibit delocalization. Consequently, a single-space description based solely on the real-space Lyapunov exponent suffers from intrinsic limitations when applied to critical states~\cite{Liu-Anomalous}.

\begin{figure}
  \centering
  \includegraphics[width=0.48\textwidth]{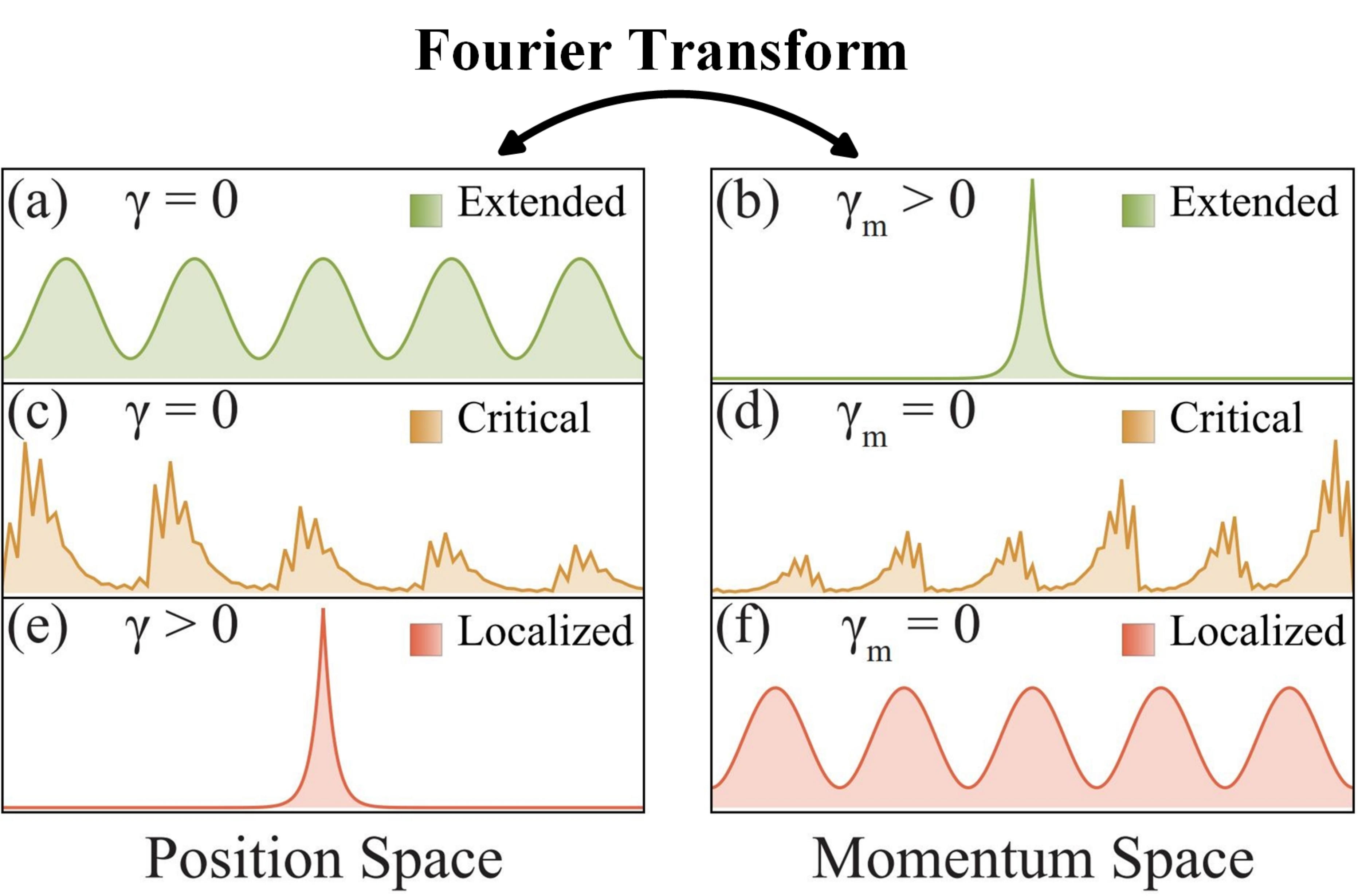}
  \caption{(Color online) Schematic illustration of typical eigenstates in position and momentum space. Panels (a), (c), and (e) show extended, critical, and localized states in position space, respectively, while panels (b), (d), and (f) display the corresponding momentum-space distributions. From the dual-space perspective, critical states are uniquely distinguished by the absence of exponential localization in both spaces. This dual absence establishes the Lyapunov exponent as a duality invariant of multifractal critical states.}
  \label{fig1}
\end{figure}

A useful conceptual perspective is provided by the Fourier uncertainty principle~\cite{MQM}, which dictates that localization in one representation necessarily implies delocalization in its Fourier-dual counterpart. Consequently, localized and extended states exhibit strongly asymmetric localization properties in position (real) and momentum spaces. This naturally raises the question of whether critical states, owing to their self-similar nature, may display a more symmetric behavior under position-momentum transformation.

Motivated by this consideration, Liu and Xia proposed a dual-space criterion for the exact identification of critical states~\cite{Liu-Critical}, overcoming the intrinsic limitations of single-representation descriptions. 
Specifically, they proposed that critical states lack exponential localization in both position and momentum space, exhibiting vanishing Lyapunov exponents and multifractal, nonergodic delocalization. 
By contrast, extended and localized states necessarily exhibit an inherent asymmetry between the two conjugate spaces: delocalization in one representation is accompanied by localization in the dual one, as illustrated in Fig.~\ref{fig1}(a), (b), (e), and (f). Critical states, however, are uniquely characterized by invariant Lyapunov exponents, indicating delocalization in both position and momentum space, as shown in Fig.~\ref{fig1}(c) and (d). \emph{This invariance of critical states under the Fourier transform, or dual-space transformation, is termed dual-space invariance.}

Dual-space invariance of critical states was first introduced by Liu and Xia in an arXiv preprint in February 2023~\cite{Liu-Critical}.
Its applicability has since been supported in a variety of systems, including flat-band lattices~\cite{quasi1} and Fibonacci topological insulators~\cite{quasi2}. 
In March 2025, Ref.~\cite{quasi3} claimed to characterize critical states through generalized incommensurate zeros in real and dual spaces, within the dual-space conceptual framework first proposed by Liu and Xia~\cite{Liu-Critical}. These developments underscore the growing interest in dual-space approaches to criticality and further highlight the foundational role of the dual-space invariance framework introduced by Liu \emph{et al.}, which predates and provides the conceptual basis for these subsequent studies.

The notion that critical states are delocalized in both position and momentum space provides an intuitive but oversimplified physical picture. This physical intuition is placed on a rigorous footing by the Liu--Xia criterion, which admits a precise mathematical formulation via the transfer-matrix method (see Appendix~\ref{app:gamma}). It identifies critical states through the simultaneous vanishing of the Lyapunov exponents in the two dual spaces,
\begin{equation}
\gamma(E,V)=\gamma_m(E,V)=0,
\end{equation}
where $\gamma$ and $\gamma_m$ correspond to position and momentum space, respectively, $E$ is the eigenenergy, and $V$ the modulation strength. 
This implies that the Lyapunov exponent serves as a duality invariant of critical states, and its advantages become evident in comparison with finite-size multifractal analysis.

According to multifractal theory~\cite{You,quasi4}, in the thermodynamic limit, the minimal scaling index $\Gamma_{\min}$ approaches $1$ for extended states and $0$ for localized states, while critical states satisfy $0<\Gamma_{\min}<1$. However, finite-size multifractal analysis suffers from intrinsic ambiguity, as critical states with small scaling exponents are difficult to distinguish from localized states.
The Liu--Xia criterion overcomes this fundamental limitation by providing an exact and representation-independent identification of critical states. This is essential because the multifractal nature of critical states cannot be uniquely characterized by a single scaling exponent, as $\Gamma_{\min}$ spans a continuous interval $(0,1)$ rather than converging to a definite value. Finally, a unified framework for Anderson localization is summarized in Table~\ref{table1}.
\begin{table}[htb]
\caption{Classification of typical eigenstates in disordered quantum systems
based on the Liu--Xia dual-space criterion~\cite{Liu-Critical}.}
\label{table1}
\begin{ruledtabular}
\begin{tabular}{c c c c c}
States & Position-Momentum & Delocalization & Multifractal & $\Gamma_{\min}$\\ \hline
Extended & $\gamma=0,~\gamma_m>0$ & Position & Absent & 1\\ 
Localized & $\gamma>0,~\gamma_m=0$  & Momentum  & Absent & 0 \\ 
Critical & $\gamma=\gamma_m=0$ & Both &  Present & $(0,1)$
\end{tabular}
\end{ruledtabular}
\end{table}

A natural question then arises: beyond Lyapunov exponents, does the dual-space characterization of critical states extend to other observables?
Here, we show that experimentally accessible quantities, particularly the inverse participation ratio, exhibit analogous scaling in position and momentum space at criticality.
This correspondence arises at the level of scaling and statistical properties rather than exact invariance.
Our results therefore complement the Liu--Xia criterion and provide a practical route to identifying multifractal critical states in both theoretical and experimental settings.

\section{Phenomenological Analysis}

Compared with the Lyapunov exponent, the inverse participation ratio (IPR) provides complementary information on wave-function localization~\cite{Longhi1,Xia,Jitomirskaya,Avila,IPR,IPR1,IPR2,IPR3}. 
It is defined for normalized wave functions as
\begin{equation}
\mathrm{IPR}=\sum_{n=1}^{L}|\psi_n|^4 ,
\end{equation}
and quantifies the spatial confinement of the probability density. 
Extended states exhibit $\mathrm{IPR}\sim L^{-1}$ in the thermodynamic limit, whereas exponentially localized states are characterized by a finite IPR. 
Related quantities, such as the participation entropy, have also been widely used to probe localization and multifractality in theoretical and experimental studies~\cite{Fan1,Fan2,Optical1,Optical2,gg0,gg1,gg2}.

For exponentially localized states, the finite IPR is consistent with a nonzero Lyapunov exponent. 
In contrast, critical states exhibit power-law decay and strong multifractal fluctuations, which preclude a description in terms of a single localization length. 
As a result, no exact analytical relation between the IPR and the Lyapunov exponent generally exists for critical states. 
Nevertheless, from a statistical perspective, the IPR is expected to display behavior analogous to that of the Lyapunov exponent.

Motivated by the Liu--Xia criterion~\cite{Liu-Critical} based on the Lyapunov exponent, we now investigate how the localization properties encoded in the IPR transform between position and momentum spaces.
Although the IPR is inherently basis dependent and therefore not invariant
under Fourier transformation (see the Appendix~\ref{app:IPR}), the dual-space invariance underlying the
Liu--Xia criterion naturally suggests that critical states should exhibit
the same scaling tendency in the two representations.
This expectation can be formulated in an operational form as
\begin{equation}
\mathrm{IPR} \sim \mathrm{IPR}_m ,
\label{eq:IPR_dual}
\end{equation}
where $\mathrm{IPR}_m$ denotes the inverse participation ratio evaluated in momentum space, and the symbol ``$\sim$'' signifies an asymptotic correspondence at the level of scaling behavior, rather than a strict equality. Equation~(\ref{eq:IPR_dual}) should therefore be interpreted as a phenomenological signature of criticality, rather than an exact invariant under Fourier transformation.

By contrast, extended and exponentially localized states typically exhibit a pronounced asymmetry between position and momentum spaces, reflected in the markedly different scaling behavior of $\mathrm{IPR}$ and $\mathrm{IPR}_m$. In comparison, critical states remain protected by dual-space symmetry and exhibit comparable localization characteristics in both representations, consistent with their multifractal and self-similar nature. On this basis, Table~\ref{table2} summarizes a phenomenological classification of eigenstates in disordered quantum systems.
\begin{table}[htb]
\caption{Phenomenological classification of eigenstates via the IPRs for finite-size systems. \label{table2}}
\begin{ruledtabular}
\begin{tabular}{c c c c c}
States & Position-Momentum & Delocalization & Multifractal & $\Gamma_{\min}$ \\ \hline
Extended & $\rm IPR\sim0,\rm IPR_m>0$ &  Position & Absent & 1 \\ 
Localized & $\rm IPR>0,\rm IPR_m\sim0$ & Momentum & Absent & 0 \\ 
Critical & $\rm IPR\sim\rm IPR_m$ &  Both & Present & $(0,1)$
\end{tabular}
\end{ruledtabular}
\end{table}

This demonstrates that the Lyapunov exponent itself is an exact duality invariant for multifractal critical states, whereas the IPR, owing to its intrinsic basis dependence, is not strictly duality-invariant. 
It should be noted that the dual-space invariance discussed here refers to the invariance of localization characteristics at the level of scaling behavior and statistical properties, rather than an exact equality of critical wave functions. In systems with self-dual symmetry, the critical states at the self-dual point remain invariant under the position-momentum transformation. In contrast, without self-dual symmetry, a critical state in position space is mapped onto a distinct critical state in momentum space, while preserving the same scaling and statistical characteristics. This demonstrates that the dual-space invariance implied by the Liu--Xia criterion~\cite{Liu-Critical} is a genuinely nontrivial result. Rather than arising from self-duality that relates identical critical states, it reveals universal features shared by distinct critical states in position and momentum spaces, establishing a fundamentally new understanding of criticality in Anderson localization.

\section{Numerical verification}
To verify previous predictions, we perform numerical simulations to investigate critical states in two representative quasiperiodic models. These systems are chosen because their Hamiltonians can be represented exactly in both position and momentum spaces, making them ideal for analysis.

\subsection{Aubry-Andr\'{e}-Harper model}
The Aubry-Andr\'{e}-Harper (AAH) model~\cite{AAH1,AAH2} represents a fundamental paradigm for studying localization phenomena in quasiperiodic systems. In position space, its Hamiltonian takes the form of a discrete Schr\"{o}dinger equation:
\begin{equation}\label{eq5}
\psi_{n + 1} + \psi_{n - 1} + V\cos (2\pi \alpha n + \theta )\psi_n = E \psi_n.
\end{equation}
where $\psi_n$ is the wave function amplitude of the particle at site $n$, $V_n=V\cos (2\pi \alpha n + \theta )$ is the quasiperiodic potential, and $E$ is the eigenvalue. $\alpha$ is an irrational number, usually taken to be the golden ratio $\alpha = \frac{{\sqrt{5} - 1}}{2}$. $\theta$ is a phase factor, typically taken to be zero.

Utilizing the Fourier transform $\psi_n=\sum_k e^{-i 2\pi \alpha n k}\phi_k$, the Hamiltonian of the AAH model in momentum space can be readily expressed,
\begin{equation}\label{eq6}
\frac{V}{2}(\phi_{k + 1} + \phi_{k - 1}) + 2\cos (2\pi \alpha k + \vartheta )\phi_k = E \phi_k.
\end{equation}

The AAH model exhibits a quantum phase transition driven by the quasiperiodic potential strength $V$. For weak quasiperiodic potential ($V<2$), all eigenstates are extended Bloch-like states, resulting in metallic transport. In contrast, for $V>2$, the system undergoes Anderson localization, where all eigenstates become exponentially localized, suppressing particle diffusion. At the critical point $V=2$, the Hamiltonians in real space [Eq.~(\ref{eq5})] and momentum space [Eq.~(\ref{eq6})] take identical forms. Consequently, the system reaches the self-dual point and exhibits multifractal eigenstates.

The procedure for identifying multifractal critical states according to the Liu--Xia criterion is outlined as follows. From Avila's global theory~\cite{Avila1,Avila2}, the Lyapunov exponents in position and momentum spaces are given by $\gamma = \ln(V/2)$ and $\gamma_m = \ln(2/V)$, respectively. Imposing the Liu--Xia criterion, $\gamma = \gamma_m$, rigorously and uniquely determines the critical point at $V = 2$, thereby establishing the multifractal nature of the eigenstates. This result is in exact agreement with that obtained from the self-duality relation, further confirming the validity and consistency of the criterion.

\subsection{Quasiperiodic-Nonlinear-Eigenproblem}
The AAH model is well-known for its critical states at the self-dual point. So, does there exist a model without self-dual symmetry that still hosts critical states over a broad parameter range?
Liu and Xia introduced such a model~\cite{Liu-Critical}, which is non-Hermitian. The construction of their model was inspired by the generalized eigenvalue problem. Mathematically, an eigenvalue problem takes the form:
\begin{equation}\label{eq7}
\hat{H} \left| \psi \right\rangle = E \hat{B} \left| \psi \right\rangle,
\end{equation}
where $\hat{H}$ denotes the Hamiltonian operator and $E$ represents the corresponding eigenvalue.
The mathematical structure of Eq.~(\ref{eq7}) exhibits a fundamental dichotomy based on the form of $\hat{B}$. When $\hat{B}=I$, the equation reduces to a conventional linear eigenvalue problem. However, when $\hat{B}$ departs from the identity matrix, the problem transforms into either a nonlinear or generalized eigenvalue problem, introducing rich mathematical complexity absent in standard quantum systems.

Thus, the Quasiperiodic-Nonlinear-Eigenproblem (QNE) model~\cite{Liu-Critical} in momentum space can be formulated as
\begin{equation}\label{eq8}
\begin{aligned}
&\{2\cos [2 \pi \alpha (k+1)]+V\}{\psi_{k+1}}+ \\
&\{2\cos [2 \pi \alpha (k-1)]-V \}\psi_{k-1}=E(\psi_{k+1}+\psi_{k-1}).
\end{aligned}
\end{equation}
Non-Hermiticity in the QNE model originates from nonlinear eigenvalue terms, fundamentally distinct from the three established paradigms of asymmetric hopping, gain/loss, and complex potentials~\cite{NH1,NH4}. This unconventional mechanism provides a new route to non-Hermitian spectra. 
As a direct consequence, the QNE model exhibits an expanded critical regime. Compared with the AAH model, where critical states exist only at the self-dual point $V = 2$, the QNE model supports critical states over a broad parameter range $0 < V \leq 2$.

After Fourier transformation, the real-space QNE Hamiltonian reduces to a linear eigenvalue problem that can be exactly diagonalized, dramatically reducing the computational cost (see Appendix~\ref{app:fourier} for details),
\begin{equation}\label{eq9}
\phi_{n + 1} + \phi_{n - 1} + i V \tan (2\pi \alpha n )\phi_n = E \phi_n.
\end{equation}
where $i$ represents the imaginary unit, and disregarding the phase factor.

\begin{figure}
  \centering
  \includegraphics[width=0.48\textwidth]{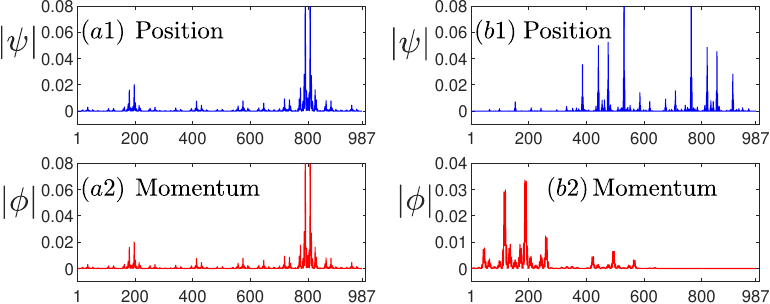}\\
  \caption{(Color online) 
  Panels (a1) and (a2) show the wave function of the AAH model ($V=2$, $E=0.0009$) in real and momentum space, respectively, while (b1) and (b2) show the corresponding results for the QNE model ($V=1$, $E=0.4$). The system size is $L=987$. Due to self-duality, critical states of the AAH model are invariant between real and momentum space. In contrast, the QNE model lacks self-duality: a critical state transforms into a distinct critical state under Fourier transformation, while their Lyapunov exponents remain identical.}
  \label{fig2}
\end{figure}

Clearly, unlike the AAH model, the QNE model lacks self-duality symmetry, and its critical states generally exhibit different structures in real and momentum spaces. Consequently, the self-duality relation completely fails to identify the critical states. In sharp contrast, the Liu--Xia criterion remains fully operative. Specifically, the Lyapunov exponents in real and momentum spaces can be rigorously obtained from Eq.~(\ref{eq9}) and Eq.~(\ref{eq8}), respectively, and the critical states are uniquely determined by enforcing their equality, as demonstrated in Ref.~\cite{Liu-Critical}. Despite the absence of any self-duality symmetry, the critical states are still dictated by the invariance of the Lyapunov exponents in dual spaces, thereby revealing the fundamental origin, conceptual advance, and universal validity of the Liu--Xia criterion.

\subsection{Verification results}
\begin{figure}
  \centering
  \includegraphics[width=0.48\textwidth]{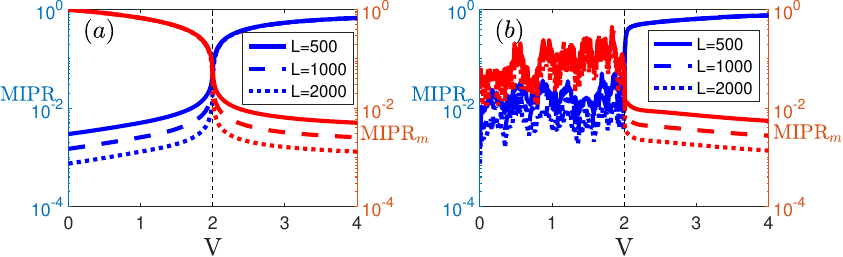}\\
\caption{
(a) AAH model: MIPR and MIPR$_m$ exhibit distinct scaling away from the critical point $V=2$, but converge to comparable values at criticality. Results are averaged over all eigenstates for $L=500,\;1000,\;2000$.
(b) QNE model: in the regime $0<V<2$, both MIPR and MIPR$_m$ display strong fluctuations yet remain comparable, supporting criticality over a broadened parameter range. Results are obtained from six eigenstates near $E=0.00001$ for $L=500,\;1000,\;2000$.}
  \label{fig3}
\end{figure}

We perform direct numerical diagonalization for systems with $L=987$ to obtain the eigenvalues and eigenstates of both the AAH and QNE models. The results are shown in Fig.~\ref{fig2}. As seen in panels (a1), (a2) and (b1), (b2), the critical states exhibit simultaneous delocalization, multifractality, and self-similarity in both position space ($|\psi|$) and momentum space ($|\phi|$). These results provide direct numerical evidence that critical wave functions obey dual-space invariance under Fourier transformation.

Figure~\ref{fig3} compares the mean inverse participation ratios in position space (MIPR) and momentum space (MIPR$_m$) for the two quasiperiodic models. The MIPR is computed by averaging the IPR over eigenstates within a narrow energy window to suppress sample fluctuations. For the AAH model, the number of sampled states scales with system size, whereas for the QNE model, the averaging is limited to a few eigenstates due to computational cost.

As shown in Fig.~\ref{fig3}(a), the AAH model exhibits strong asymmetry between position and momentum space away from criticality. For $V<2$, the real-space MIPR decreases with system size, consistent with extended states, while the momentum-space MIPR$_m$ remains finite, indicating localization in momentum space. The opposite behavior occurs for $V>2$. At the self-dual point $V=2$, the two quantities become comparable and follow similar scaling, reflecting dual-space invariance at criticality.

The behavior of the QNE model, shown in Fig.~\ref{fig3}(b), is qualitatively different. For $V>2$, the eigenstates are exponentially localized in real space, leading to a weak size dependence of the MIPR, while the momentum-space MIPR$_m$ exhibits pronounced system-size scaling. In contrast, for $0<V<2$, both MIPR and MIPR$_m$ show strong fluctuations and nontrivial size dependence. Rather than approaching plateaus characteristic of extended or localized phases, the two quantities exhibit comparable scaling, consistent with a critical regime.

We emphasize that, due to finite-size effects and the basis dependence of the IPR, exact equality between MIPR and MIPR$_m$ is neither expected nor required. Instead, the results show that critical states are characterized by comparable scaling in position and momentum space. These findings provide direct numerical support for the dual-space correspondence.

\section{Experimental feasibility}

The dual-space correspondence identified in this work is well suited for
experimental verification in ultracold atomic systems, where both
position- and momentum-space observables are directly accessible.
One-dimensional quasiperiodic lattices realizing AAH-type Hamiltonians can
be implemented using optical lattices with superimposed incommensurate
potentials.

In such setups, the real-space density distribution $|\psi_n|^2$ can be
measured using in situ fluorescence or absorption imaging~\cite{gg9}, allowing direct
evaluation of the real-space IPR.
The momentum-space distribution $|\phi_k|^2$ can be obtained via standard
time-of-flight (TOF) measurements~\cite{Biddle}. After switching off the trapping and
lattice potentials, the atomic cloud undergoes ballistic expansion, and in
the long-TOF limit the measured density profile reflects the initial
momentum distribution. By discretizing the TOF images according to the
imaging resolution, the momentum-space inverse participation ratio
$\mathrm{IPR}_m$ can be extracted in a directly analogous manner.

Within this framework, extended and localized states exhibit pronounced
asymmetry between position and momentum space, whereas critical states show
comparable localization properties in both representations. Although finite
system sizes and experimental resolution prevent exact quantitative
agreement, the scaling-level correspondence predicted here remains observable
under realistic conditions. These results demonstrate that current cold-atom
experiments provide a feasible and practical route for identifying
multifractal critical states through dual-space diagnostics.

\section{Summary and outlook}

In summary, we have investigated critical states in one-dimensional
quasiperiodic systems from a dual-space perspective.
Using analytical arguments and numerical simulations of representative
models, including the AAH model and a quasiperiodic nonlinear eigenproblem,
we have shown that critical states are uniquely characterized by
comparable scaling in dual spaces.
This property sharply distinguishes them from extended and localized
states, which exhibit a pronounced asymmetry between position and momentum
spaces.

We have further demonstrated that the inverse participation ratio provides
a simple and practical diagnostic of this behavior: critical states exhibit
comparable scaling of the inverse participation ratio in position and
momentum space even in finite systems.
This result establishes dual-space correspondence as an operational
criterion for identifying multifractal criticality beyond conventional
approaches based solely on a single space.

An important open direction is to examine interacting many-body systems,
where quasiperiodicity and interactions can produce many-body localized and
critical phases.
It will be particularly interesting to determine whether a similar scaling
correspondence emerges between real-space and Fock-space or momentum-space
structures of many-body eigenstates, and whether the Liu--Xia criterion can be extended to provide a precise characterization of many-body criticality.

\begin{acknowledgments}
This work was supported by the Natural Science Foundation of Nanjing University of Posts and Telecommunications (Grants No. NY223109).
\end{acknowledgments}


\begin{appendix}
\section{Computation of the Lyapunov exponents $\gamma$ and $\gamma_m$
via transfer-matrix methods}
\label{app:gamma}

\subsection*{1. Real-space Lyapunov exponent $\gamma$}

For a given energy $E$, the real-space Lyapunov exponent $\gamma(E)$
characterizes the asymptotic exponential growth of the wave-function
amplitude along the lattice.
It is computed using the standard transfer-matrix method.
Starting from an initial vector $\mathbf{v}_0$ (e.g., $(1,0)^{\mathrm T}$),
one iterates the site-dependent transfer matrix $\mathbf{T}_n(E)$,
\begin{equation}
\mathbf{v}_{n+1}=\mathbf{T}_n(E)\mathbf{v}_n.
\end{equation}
After $N$ steps, the total transfer matrix is
$\boldsymbol{\Pi}_N(E)=\mathbf{T}_{N-1}\cdots\mathbf{T}_0$, and the
Lyapunov exponent is defined as
\begin{equation}
\gamma(E)=\lim_{N\to\infty}\frac{1}{N}
\ln\left\|\boldsymbol{\Pi}_N(E)\right\|.
\label{eq:gamma_real}
\end{equation}
In numerical implementation, periodic renormalization is applied to prevent
overflow, and $\gamma(E)$ is obtained from the averaged logarithmic growth
rate for large $N$ (typically $N\sim10^5$).

\subsection*{2. Momentum-space Lyapunov exponent $\gamma_m$}

The momentum-space Lyapunov exponent $\gamma_m(E)$ is computed using an
analogous transfer-matrix formulation in momentum space.
After Fourier transforming the eigenvalue equation, the momentum-space
amplitudes $\tilde{\psi}_k$ satisfy a linear recurrence relation.
For quasiperiodic potentials with a single incommensurate frequency, this
relation forms an effective one-dimensional chain,
\begin{equation}
a_k \tilde{\psi}_{k+Q}
+b_k \tilde{\psi}_k
+c_k \tilde{\psi}_{k-Q}
=E\tilde{\psi}_k,
\end{equation}
which can be rewritten as
\begin{equation}
\begin{pmatrix}
\tilde{\psi}_{k+Q}\\
\tilde{\psi}_k
\end{pmatrix}
=
\mathbf{M}_k(E)
\begin{pmatrix}
\tilde{\psi}_k\\
\tilde{\psi}_{k-Q}
\end{pmatrix}.
\end{equation}
Iterating the corresponding transfer matrices yields
\begin{equation}
\gamma_m(E)=\lim_{K\to\infty}\frac{1}{K}
\ln\left\|\boldsymbol{\Pi}^{(m)}_K(E)\right\|,
\end{equation}
where $\boldsymbol{\Pi}^{(m)}_K(E)$ denotes the product of momentum-space
transfer matrices.

In both cases, the Lyapunov exponent quantifies the exponential growth rate
of the transfer-matrix norm, with the only difference being whether the
recurrence is defined in position or momentum space.

\section{Contrast Between Basis-Dependent IPR and Basis-Invariant Lyapunov Exponents}
\label{app:IPR}
It is instructive to contrast the distinct roles of the Lyapunov exponent and the inverse participation ratio (IPR) in characterizing localization. The Lyapunov exponent is defined by the asymptotic exponential growth or decay of solutions to a linear recurrence relation and thus represents an intrinsic dynamical quantity associated with the transfer-matrix structure. Consequently, it is independent of the choice of basis and depends only on the underlying recursive dynamics in a given representation (real or momentum space).

In contrast, the IPR depends explicitly on the expansion coefficients of the wave function in a chosen basis and quantifies the degree of localization within that representation. Because higher-order moments of the wave-function amplitudes are not invariant under unitary transformations, the IPR is inherently basis dependent and generally differs between position and momentum space.

This distinction is central to the present work. While the simultaneous vanishing of Lyapunov exponents in dual spaces provides a basis-invariant criterion for criticality, the emergence of comparable scaling of the IPR in dual spaces constitutes a nontrivial physical signature. It indicates that, despite its basis dependence, the IPR reflects the reduced asymmetry between position and momentum space at criticality.

\section{Fourier transformation for the QNE model}
\label{app:fourier}

We show that the real-space equation
\begin{equation}
\phi_{n+1}+\phi_{n-1}
+iV\tan(2\pi\alpha n)\phi_n
=E\phi_n
\label{eq:qne_real}
\end{equation}
transforms into the momentum-space equation
\begin{equation}
\begin{aligned}
&\bigl[2\cos\bigl(2\pi\alpha(k+1)\bigr)+V\bigr]\psi_{k+1}
\\
&\quad+
\bigl[2\cos\bigl(2\pi\alpha(k-1)\bigr)-V\bigr]\psi_{k-1}
=
E(\psi_{k+1}+\psi_{k-1}),
\end{aligned}
\label{eq:qne_momentum}
\end{equation}
under a discrete Fourier transform.

\subsection*{1. Fourier transform}

Multiplying Eq.~\eqref{eq:qne_real} by $\cos(2\pi\alpha n)$ gives
\begin{equation}
\begin{aligned}
&\cos(2\pi\alpha n)\phi_{n+1}
+\cos(2\pi\alpha n)\phi_{n-1}
\\
&\quad
+iV\sin(2\pi\alpha n)\phi_n
=
E\cos(2\pi\alpha n)\phi_n .
\end{aligned}
\label{eq:qne_multiplied}
\end{equation}

We define the discrete Fourier transform
\begin{equation}
\psi_k=\sum_n e^{-i2\pi\alpha kn}\phi_n, \qquad
\phi_n=\sum_k e^{i2\pi\alpha kn}\psi_k .
\end{equation}

\subsection*{2. Momentum-space equation}

Using standard trigonometric identities, the Fourier transform yields
\begin{align}
\sum_n e^{-i2\pi\alpha kn}
\cos(2\pi\alpha n)\phi_{n\pm1}
&=
\cos\bigl[2\pi\alpha(k\pm1)\bigr]\psi_{k\pm1},
\\
\sum_n e^{-i2\pi\alpha kn}
\sin(2\pi\alpha n)\phi_n
&=
\frac{1}{2i}\bigl(\psi_{k-1}-\psi_{k+1}\bigr),
\\
\sum_n e^{-i2\pi\alpha kn}
\cos(2\pi\alpha n)\phi_n
&=
\frac12\bigl(\psi_{k-1}+\psi_{k+1}\bigr).
\end{align}

Substituting into Eq.~\eqref{eq:qne_multiplied} gives
\begin{equation}
\begin{aligned}
&\cos\bigl[2\pi\alpha(k+1)\bigr]\psi_{k+1}
+\cos\bigl[2\pi\alpha(k-1)\bigr]\psi_{k-1}
\\
&\quad
+\frac{V}{2}(\psi_{k-1}-\psi_{k+1})
=
\frac{E}{2}(\psi_{k-1}+\psi_{k+1}),
\end{aligned}
\end{equation}
which reduces to Eq.~\eqref{eq:qne_momentum}.

\end{appendix}

\end{document}